\documentclass{pasj00}
\draft

\begin{document}
\SetRunningHead{Author(s) in page-head}{Running Head}
\Received{2002/10/24}
\Accepted{2002/11/19}

\title{Evidence for Jet Collimation in SS~433 \\ with the Chandra HETGS}

\author{
  Masaaki \textsc{Namiki},\altaffilmark{1,2} 
  Nobuyuki \textsc{Kawai},\altaffilmark{1,3}  
  Taro \textsc{Kotani},\altaffilmark{3,4}   and  
  Kazuo \textsc{Makishima}\altaffilmark{1,5}}
\altaffiltext{1}{The Institute of Physical and Chemical Research (RIKEN), \\
2-1, Hirosawa, Wako, Saitama. 351-0198}
\altaffiltext{2}{Department of Physics, Tokyo University of Science, \\
1-3 Kagurazaka, Shinjuku-ku, Tokyo. 162-8601}
\email{namiki@crab.riken.go.jp}
\altaffiltext{3}{Department of Physics, Tokyo Institute of Technology, \\
2-12-1, Ookayama, Meguro, Tokyo. 152-0033}
\email{nkawai@tithp1.hp.phys.titech.ac.jp}
\altaffiltext{4}{Laboratory for High Energy Astrophysics, NASA Goddard Space Flight Center, \\
Greenbelt, MD, 20771, U.S.A.}
\email{kotani@milkyway.gsfc.nasa.gov}
\altaffiltext{5}{Department of Physics, University of Tokyo, \\
7-3-1, Hongo, Bunkyo-ku, Tokyo. 113-0033}
\email{maxima@phys.s.u-tokyo.ac.jp}


\maketitle


\begin{abstract}

High-resolution X-ray spectra of SS~433 obtained after a binary egress 
with the Chandra High Energy Transmission Grating Spectrometer (HETGS) were studied. 
Many Doppler-shifted X-ray emission lines from highly ionized elements were   
detected.  The initial temperature of the jets is estimated to be 20 keV. 
The lines are found to generally be broader than the instrumented resolution. 
The widths of the Fe {\sc xxv} K$\alpha$ and Si {\sc xiii} K$\alpha$ lines 
correspond to velocity dispersions of 2100 $^{+600}_{-340}$ km s$^{-1}$ and 840 
$^{+180}_{-150}$  km s$^{-1}$ respectively, in terms of Gaussian sigma. 
Neither the measured line widths nor their dependence on the atomic number 
can be explained by thermal broadening alone.  
Alternative explanations of the observed line widths are discussed, 
including in particular a progressive jet collimation along its axis. 

\end{abstract}

\section{Introduction}

SS~433 is an enigmatic X-ray binary (orbital period $\sim$13.1 d) 
with bipolar jets ejecting matter at a relativistic velocity of 0.26$-$times 
the light speed. The jet axis precesses with a period of $\sim$ 162.5 d.  
Although SS~433 has been studied for more than 20 years since its discovery,  
its fundamental properties, such as the origin of the jet's acceleration  
and precession, and the nature of the compact object, remain unknown.  

The first intensive X-ray studies of SS~433 were performed by the Einstein Satellite.  
Then, SS~433 was revealed to be a variable X-ray source 
(Grindlay et al.  1984;  Band, Grindlay 1986). 
Indeed, Watson et al.  (1986)  and  Matsuoka et al.  (1986) 
soon after found a Doppler-shifted  Fe{\sc xxv} K$\alpha$ emission line in 
the X-ray spectra of SS~433 obtained by  EXOSAT and the Tenma, respectively; 
they detected one emission line, and Watson et al.  (1986) 
concluded that the X-ray emitting 
part of the receding jet is hidden behind the accretion disk.  
Thus, the length of the X-ray jet was estimated to be shorter than the orbital 
separation, $\stackrel < \sim 10^{12}$ cm.  
From the Ginga data, Kawai et al.  (1989) determined the temperature 
of the jet to be 35 keV by modeling the continuum with  thermal bremsstrahlung.  
Brinkmann et al.  (1991) applied a hydrodynamical numerical model to the 
 Ginga data, and estimated the initial temperature of the jet to be as 
high as $\sim$ 40--70 keV.  Since their solution implied a short 
($\sim 10^{10}$ cm) jet with a high ($\sim 10^{13}$ 
cm$^{-3}$) density, the receding jet was considered to be easily hidden by 
the accretion disk, to become consistent with the  EXOSAT data.  

In 1993, however, a spectrum taken with  ASCA, with its high-energy 
resolution and high sensitivity  has overthrown the view so far established: 
the spectrum showed full ``pairs'' of Doppler-shifted  emission lines from 
heavy elements such as 
Si {\sc xiv}, S {\sc xv}, S {\sc xvi}, Ar {\sc xvii}, Ar {\sc xviii}, 
Ca {\sc xix}, Fe {\sc xxv}, Fe {\sc xxvi}, and Ni {\sc xxvii} 
(Kotani et al. 1994; Kawai 1995).  
Their detection implies that the approaching and receding jets are both 
visible, so that the X-ray jets are longer ($\sim$ 10$^{13}$ cm) than 
previously estimated. Kotani et al.  (1996)  determined the base 
temperature of the jet to be 20 keV from the line flux ratio of Fe {\sc xxvi} 
K$\alpha$ to Fe {\sc xxv} K$\alpha$. Further analysis of the spectra revealed 
that the kinetic luminosity of SS~433 far exceeds the Eddington luminosity for 
any stellar size object (Kotani 1998). Kotani (1998) also suggested that 
the highly ionized iron K$\alpha$ lines are clearly seen in all of the data 
sets, while the low-energy emission lines from the receding jet are  
generally weaker than those from the approaching jet, and sometimes absent. 
This can be explained if absorbing matter is present in the line of sight 
to the receding jet. 

Although the ASCA spectra thus revealed many emission lines which no 
previous mission was able to detect,  the energy resolution of the 
ASCA SIS is still insufficient to fully resolve the emission line 
forest; this is particularly the case at energies below $\sim$ 3 keV, where 
lines from different elements overlap and the effect of the absorption is 
severest. The Chandra High Energy Transmission Grating Spectrometer (HETGS) is 
ideal for an attempt to significantly improve the knowledge obtained with  
ASCA. An important feature of the HETGS is that spectra are obtained 
independently and simultaneously using the High Energy Gratings (HEGs) and 
Medium Energy Gratings (MEGs).  The HEGs have a twice-higher spectral 
resolution and a larger effective area above 4 keV than the MEGs, while the 
MEGs have an extended bandpass and a larger effective area for detecting 
low-energy emission lines.  The energy resolution (FWHM) of the HEGs is 
$\sim$ 0.2\% at 2.0 keV and $\sim$ 0.5\% at 6.0 keV, while that of MEGs is 
$\sim$ 0.14\% at 0.8 keV and $\sim$ 0.33\% at 2.0 keV.  Marshall et al.  (2002) 
reported on the first observation of SS~433 using the Chandra/HETGS, made on 
1999 September 23. The lines were measurably broadened to 1700 km s$^{-1}$ 
(FWHM), and the widths did not depend significantly on the line-center energy, 
suggesting that the emission occurs in a freely expanding region of constant 
collimation with an opening angle of 1$^{\circ}$.26 $\pm$ 0$^{\circ}$.06.  
Furthermore, the excellent angular resolution of Chandra has revealed for the 
first time that the image of SS~433 is extended along the east-west direction 
on a scale of 2$''$--5$''$, which is comparable to that observed in the 
radio band (Hjellming, Johnston 1981). These impressive results suggest 
that the Chandra X-ray Observatory will give us another surprise regarding this 
fascinating source. 

\section{Observations}\label{observation}

The present observation of SS~433 was performed with the Chandra HETGS 
on 2001 May 12 UT10:38--16:33, 
which was on schedule with our proposal. 
The total exposure time was 19.7 ks.
The corresponding orbital phase was 0.2832 $\pm$ 0.0094 (Gladyshev et al.  1987), when 
the compact object (including the two jets) and the accretion disk emerge out of an eclipse 
by the companion star. 
The Doppler-shift parameters at this observation were expected to be around 
$z_{\rm bl}\sim$ $+$0.075  and $z_{\rm rd}\sim$ $-$0.004  for the 
approaching and receding jets, respectively, according to the ephemeris of  Margon and Anderson (1989).  
Usually, the jet directed to the east (called ``blue jet'') is approaching us, 
and that to the west (``red jet'') is receding. 
However, the expected precessional phase on 2001 May 12 was reversed in such a way that 
the ``blue jet'' is receding and the ``red jet'' is approaching.  
This reversal occurs for a minor fraction ($\sim$ 50 d) of the 
162.5 d precession period. 

The observation log is shown in table~1.  
All the data were acquired in the ACIS-S faint mode with 
CCD chips S0, S1, S2, S3, S4 and S5, and processed in a standard way. 
The average flux of SS~433 in the 1--10 keV band was 
$8.7 \times 10^{-11}$ erg s$^{-1}$ cm$^{-2}$, 
corresponding to a luminosity of $2.4 \times 10^{35}$ erg s$^{-1}$, assuming a 
source distance of 4.85 kpc (Vermeulen et al. 1993). 
This luminosity is typical of SS~433. 

\section{Analysis and Results}\label{analysis}

\subsection{Data Reduction}\label{reduction}

Figure~1 shows the zeroth-order image and the dispersion lines of 
SS~433 taken with the ACIS-CCD.  
Each of the ``arms'' of $X$ pattern, formed by diffraction, yields the  
first-order spectrum specified by the grating type (HEG or MEG) and sign of the order (plus or minus).  
In the image, there are no obvious point sources, except for SS~433. 
On-source event counts have been accumulated on the dispersed image over narrow 
strips along the dispersion lines, of width 2$''$.39 on either side. 
Background counts have been accumulated along similar strips, 
on both sides of each dispersion line, with a distance range of 2$''$.39--23$''$.9. 
The present study utilizes the dispersed data of $+$1 and $-$1 orders, by co-adding them 
together because no significant discrepancy is seen between them. 

The spectral extraction and data reduction were all performed with 
the standard pipeline for the Chandra HETGS provided by the 
Chandra X-ray Center (CXO). 
The following data analysis utilizes the Chandra Interactive Analysis of Observations 
(CIAO) version 2.0 and custom routines in XSPEC. 
The spectra obtained by the HEG/HETGS in the total energy band are 
shown in figure~2. 

When the energy-bin widths are chosen to be comparable to the energy 
resolution of the data, the statistics for each energy bin are so limited 
that the  $\chi ^2$ minimization criterion is inappropriate for evaluating the spectral model 
fits. Instead, the fits are evaluated via minimum likelihood method, using the function 
\begin{eqnarray*}
C = 2 \sum ^{N} _{i=1} [ y(x_{i}) - y_{i} + y_{i}({\rm ln} y_{i} - {\rm ln} y(x_{i}))], 
\end{eqnarray*}
where $x_{i}$ is the energy at the $i$-th bin, 
$y_{i}$ is the observed data at $x_{i}$, and $y(x_{i})$ is the value of the fitting function.  
Minimizing $C$ gives the best-fit model parameters (Cash 1979; 
XSPEC User's Guide for version 11.2.x, p.151).  
The $C$-statistic assumes that the errors on 
the counts are purely Poissonian, and hence it cannot deal with 
background-subtracted data. 
Therefore, the background is not subtracted in the following analysis. 
This is justified, because the background count rate is only 0.4\% of the signal count rate, 
as is clear from figure~2. 

The jet precession parameters of Margon and Anderson (1989)  predict the Fe {\sc xxv} K$\alpha$ 
lines from the blue and red jets to appear at energies of 6.236 and 6.718 keV, respectively. 
Indeed, the on-source spectrum exhibits a strong feature over an energy range of 6.3--6.8 keV 
(figure~2). 
In an expanded spectrum over this energy range, given in figure~3, 
the emission feature is resolved into a series of emission lines. However, none of them corresponds 
in energy to the predicted Fe {\sc xxv} K$\alpha$ lines. 
This suggests that the jet-precession phase has deviated from the prediction.  
Actually, a series of optical spectroscopy of SS~433, performed at Gunma Astronomical Observatory (GAO)
over a period spanning from 2001 May 12 (simultaneous with the present Chandra observation) to the end 
of 2001 November, consistently imply $\sim$ 20 d phase advance from the prediction 
(a private communication with K.~Kinugasa and H.~Kawakita; Namiki 2003 in preparation). 
Below, the jet redshifts are re-evaluated in reference to the optical data. 

\subsection{Spectral Fitting in the Hard Band}\label{hardspec}

In figure~3, the HEG/HETGS spectrum in 5.5--8.5 keV energy band around 
the iron lines is fitted with a model consisting of 
a power-law continuum and narrow Gaussian lines. 
The fitting model is given, as a function of energy $E$, as
\begin{eqnarray}
&&e^{-\sigma(E)N_{\rm H}} \times \left [ A \times E^{-\mit\Gamma} 
+\mbox{Fe {\sc i} K}\alpha_{z = 0} 
+\mbox{Fe {\sc i} K}\beta_{z = 0} \right .\nonumber\\
&&+ \left ( 
  \mbox{Fe {\sc xxv}   K}\alpha
+ \mbox{Fe {\sc xxvi}  K}\alpha
+ \mbox{Ni {\sc xxvii} K}\alpha
+ \mbox{Fe {\sc xxv}   K}\beta \right ) _{z = z_{\rm red}}\\
&&\left . + \left (
  \mbox{Fe {\sc xxv}   K}\alpha
+ \mbox{Fe {\sc xxvi}  K}\alpha
+ \mbox{Ni {\sc xxvii} K}\alpha
+ \mbox{Fe {\sc xxv}   K}\beta \right ) _{z = z_{\rm blue}}\right ]\nonumber, 
\end{eqnarray}
where $\sigma (E)$ is the photo-electric cross-section
at $E$, $N_{\rm H}$ is the equivalent hydrogen column density, while 
$A$ and $\mit\Gamma$ are the normalization and photon index of the power-law continuum, 
respectively. The fitting model includes Gaussian components corresponding to 
ten emission lines, as specified by the line names in equation (1): 
four pairs of Doppler-shifted ionized lines 
(Fe {\sc xxv} K$\alpha$, Fe {\sc xxvi} K$\alpha$, Ni {\sc xxvii} K$\alpha$ and Fe {\sc xxv} K$\beta$) 
from the two jets, and two stationary lines (Fe {\sc i} K$\alpha$ and Fe {\sc i} K$\beta$). 
The rest-frame center energies of the lines were fixed to the 
theoretical values, while the overall Doppler shifts of the blue and red jets were set free; 
the redshift is common among all of the lines belonging to the same jet. 
The model is subjected to photoelectric absorption, with $N_{\rm H}$  
fixed to the value which is determined in the lower energy band (subsection 3.3).  

On the same May 12, the Doppler shifts in the optical band were obtained using 
H$\alpha$ lines at GAO as $+$0.0424 $\pm$ 0.0076  and $+$0.0204 $\pm$ 0.0147  
for the blue and red jet, respectively. 
Considering the general agreement between the X-ray and optical Doppler shifts 
(within $\leq$ 0.64 d in the precessional phase; Kotani 1998),  
the optical redshifts  were employed as initial values of the X-ray spectral fitting. 
As a result, the energies of Fe {\sc xxv} K$\alpha$ lines were obtained at 6.410 and 6.524 keV 
($z$ = 0.0449 and 0.0267) for the blue and red jets, respectively, as given in 
table~2 and presented in figure~3. 
These values agree, within the respective errors, with the optical measurements at GAO. 
The observed precessional phase is such that the Fe {\sc xxv} K$\alpha$ lines from the two 
jets and the fluorescent iron line from the accretion disk are heavily blended together 
around 6.5 keV. 
However, thanks to the excellent energy resolution of the HETGS,  
the lines were clearly resolved, except for an overlap between Fe {\sc i} (6.399 keV) and Fe {\sc xxv} 
K$\alpha$ of the blue jet (6.410 keV); 
these two lines, both known to be generally intense, could not be separated  even with 
the energy resolution of the Chandra HETGS. 
However, this does not affect the determination of the jet redshifts, since the other lines contribute. 
Furthermore, as described later, the fluxes of the two overlapping lines can be determined separately 
with a sufficient accuracy by referring to their line widths. 

The lower panel of figure~3 
shows the ratio of the spectrum against the best-fit model 
described by equation (1), with the Gaussians all constrained to be narrow. 
There, large residuals are seen on  both sides of each iron emission line. 
Thus, the lines are inferred to be broad, as already reported by Marshall et al.  (2002) 
based on the same instrumentation. 
As shown in figure~4,  
a much better fit was obtained by 
allowing the Gaussians for the Doppler-shifted lines to have a common finite width; 
the $C$-statistics value decreased significantly from 851.94 (d.o.f. = 595) to 703.57 (d.o.f. = 594), 
and the residual between the fitting model and the data points disappeared.   
By using Monte Carlo methods to evaluate the absolute fit goodness, 
the narrow Gaussian model was completely rejected, while 
the broad Gaussian model  was acceptable at a 2.7 $\sigma$ confidence interval. 
The Gaussians for the stationary fluorescent lines were kept narrow, 
because they are thought to come from a vicinity of the compact object. 
The width (standard deviation) of the Doppler-shifted lines has been constrained as 
$\sigma$ $\sim$  39.5 $^{+7.7}_{-6.5}$  eV, 
or 1850 $^{+360}_{-300}$ km s$^{-1}$ in terms of the Doppler velocity dispersion. 
Although the Fe {\sc xxv} K$\alpha$ line is intrinsically broad due to the triplet 
structure (resonance, intercombination and forbidden), the Gaussian width 
decreases insignificantly to 35.9 $^{+7.9}_{-5.8}$  eV even considering the triplet.  
Here, the intensity ratio of the triplet lines was assumed to be 1.0 : 0.28 : 0.23 for 
the resonance, intercombination and forbidden lines, respectively,  calculated by 
Mewe et al. (1985) at a plasma temperature of 20 keV. 

The jet redshifts of SS~433 change continuously with time.  
Therefore, the line width may be caused by drifts in the redshifts during the observation. 
However, the total data span of the present observation, only 21.3 ks, is expected to cause changes 
in the iron line energies by no more than 4.0 eV, which is too small to explain the observed line 
width. For a further confirmation, the spectrum was divided into two parts, 
the former and latter half of the exposure time, 
and each data were fitted independently. 
In the case of Fe {\sc xxv} K$\alpha$, the differences of the line center energy were 
obtained as 9.0 $\pm$ 29.4 eV and 30.0 $\pm$ 41.0 eV for the blue and red jet, respectively. 
As a result, no significant difference was found  between the two spectra, and they both 
exhibited broad lines. 

\subsection{Spectral Fitting in the Soft Band}\label{softspec}

Figure~5  
shows the spectra obtained by the MEG and HEG in the soft band (1.5--4.0 keV). 
Below, they  are fitted jointly. 

First, the data were fitted with a model of a power-law continuum and narrow Gaussian lines with 
photoelectric absorption, 
just in the same way  as was performed in subsection 3.2,  
using the redshifts derived there as initial values. 
The fitting model is given as
\begin{eqnarray}
&&e^{-\sigma(E)N_{\rm H}} \times \left [ A \times E^{-\mit\Gamma} \right .\nonumber\\
&&+ \left ( 
  \mbox{Si {\sc xiii} K}\alpha
+ \mbox{Si {\sc xiv}  K}\alpha
+ \mbox{Si {\sc xiv}  K}\beta
+ \mbox{S  {\sc xv}   K}\alpha
+ \mbox{S  {\sc xvi}  K}\alpha \right ) _{z = z_{\rm red}}\\
&&\left . + \left (
  \mbox{Si {\sc xiii} K}\alpha
+ \mbox{Si {\sc xiv}  K}\alpha
+ \mbox{Si {\sc xiv}  K}\beta
+ \mbox{S  {\sc xv}   K}\alpha
+ \mbox{S  {\sc xvi}  K}\alpha \right ) _{z = z_{\rm blue}}\right ]\nonumber, 
\end{eqnarray}
where the helium-like emission lines include the resonance and forbidden components, 
and the other annotations of this model are the same as those in subsection 3.2.  
As a result, a clear local minimum in the fit $C$-statistics has been obtained at $z_{\rm bl}$ = 0.0446 and 
$z_{\rm rd}$ = 0.0220, both close to the input initial values. 
As shown in figure~5,  this fit implies the presence of 
Si {\sc xiii} K$\alpha$, Si {\sc xiv} K$\alpha$, S {\sc xv} K$\alpha$, and S {\sc xvi} K$\alpha$ lines from 
the two jet components. 

Although the spectra were thus represented approximately by equation (2), significant fit residuals 
are again observed on both sides of each line. 
As shown in figure~6, the fit was greatly improved by allowing the lines to have a finite 
common width;   
the $C$-statistics value has decreased from 1508.82 (d.o.f. = 1236) to 1468.32 (d.o.f. = 1235).  
The Monte Carlo calculation indicates that the former is completely rejected, while the latter 
is acceptable at a 5.0 $\sigma$ confidence interval. 
The model parameters were obtained as given in table~3. 
The redshift of the blue jet is consistent within the 90\% statistical error with 
that obtained in the  hard band (table~2).  
Although the redshift of the red jet is inconsistent with that in the hard band, 
it can be understood by considering the systematic errors 
($\pm$ 0.006 \AA \  for HEG;  $\pm$ 0.011 \AA \  for MEG). 
The Gaussian width (standard deviation) was obtained to be $\sigma$ = 5.2 $^{+1.1}_{-0.9}$ eV, 
or 870 $^{+190}_{-150}$ km s$^{-1}$ in terms of the Doppler velocity dispersion. 

Even in the broad-line fit in figure~6, there still remain residuals 
around 1.85 keV, approximately exhibiting a double-peaked feature. 
This feature was successfully modeled by two Gaussians with central energies 
at 1.847 and 1.874 keV and a finite width of $\sigma$ = 6.8 $^{+1.6} _{-1.2}$ eV.
Adding these two lines made the goodness-of-fit of the present model 
satisfactory at a 1.0 $\sigma$ confidence interval. 
However, taking the obtained Doppler shifts into account, 
the energies of these two additional lines do not correspond to the atomic lines of any major element, 
and hence they cannot be represented by the jet components. 

\subsection{Spectral Fitting over the Total Band}\label{bothspec}

By analyzing the HETGS spectra in the Fe K$\alpha$ and Si K$\alpha$ bands separately,  
it has been suggested that a common set of $z_{\rm bl}$ and $z_{\rm rd}$ can reproduce the 
overall data, but the Fe K$\alpha$ lines are more broadened 
(in terms of the Doppler velocity dispersion) than the Si K$\alpha$ lines. 
In order to confirm these two inferences, it is appropriate to fit the two spectral 
regions  simultaneously. 

The data were thus fitted over the total (1.5--9.0 keV) energy band 
with a model which combines those presented in equation (1) and equation (2). 
The photoelectric absorption, the photon index and intensity of the power-law, 
and the jet redshifts were constrained to be common between the two energy bands, 
while the line widths were allowed to take separate values in the hard and soft bands.   
The unidentified features (at 1.847 and 1.874 keV) were left unmodeled in order to 
reduce the number of free parameters. 
The results of subsection 3.2 and 3.3 were used for the initial parameters 
in the respective energy bands. 
The joint fit has been acceptable (at 5.0 $\sigma$), and the jet redshifts, 
line intensities, and the line widths turned out to be 
consistent with those from the individual fits. 
Although the power-law parameters and the absorption column density slightly disagree with 
those obtained in the limited energy bands, this is a relatively minor effect (table~4). 

The most important information obtained from this fit is 
the relation between the widths of low-energy and high-energy lines.  
As presented in figure~7, the results clearly confirm that 
the velocity dispersion of the high energy lines is 
significantly larger than that of the low-energy lines.

\section{Discussion}\label{discuss}

\subsection{Jet Temperature}\label{flux}

The intensity ratios among  emission lines provide important information on the physical 
conditions of cosmic hot plasmas. In particular, the 
intensity ratio of the K$\alpha$ line from Fe {\sc xxvi} to that of Fe {\sc xxv} 
provides a clear temperature indicator for very hot plasmas, and hence 
can be used to constrain the jet parameters of SS~433.  
By comparing the  ASCA measurements with detailed model calculations, 
Kotani et al.  (1996) estimated the typical initial temperature of the jets 
to be  $T_0$ $\sim$ 20 keV.  
In the present observation, 
the intensity ratio of the K$\alpha$ line from Fe {\sc xxvi} to that of Fe {\sc xxv} was 
obtained as 0.46 $\pm$ 0.17 and 0.42 $\pm$ 0.17, for the blue and red jet, respectively, 
which translate to $T_0$ = $19.5^{+8.8}_{-6.9}$ and $T_0$ = $17.2^{+8.0}_{-6.7}$ keV 
according to the model of Kotani et al.  (1996).  
From these values, the jet activity of SS~433 is inferred to have been 
in a normal state during the observation. 

In contrast, Marshall et al.  (2002) derived significantly lower values of $T_{0}$, 
based on the Chandra HETGS observation on 1999 September 23 (orbital phase $\sim$ 0.67). 
In the spectra of Marshall et al.  (2002), the Fe {\sc xxvi} lines from the jets 
were so weak that the intensity ratios of the iron lines were  
0.30 $\pm$ 0.12 and 0.17 $\pm$ 0.08 for the blue and red jets, respectively, 
yielding $T_0$ = 12.9$^{+4.7}_{-4.3}$ and $T_0$ = 8.6 $^{+2.4}_{-2.1}$ keV.  
Because these temperatures are lower than the usually reported values ($\sim$ 20 keV),  
it appears that the high-energy lines, emitted near the base of the jet of SS~433, 
were at that time hidden by something, such as an expanded accretion disk. 
Thus, it is likely that Marshall et al.  (2002) observed an unusual low-activity  
state of SS~433. 
The present paper provides the first report on the high-resolution X-ray spectroscopy 
of SS~433 in a relatively normal state, which is close to the  past usual states 
observed with Ginga and ASCA. 

\subsection{Line Widths}\label{width}

The Doppler-shifted lines of SS~433, measured with the HETGS, 
have been successfully represented by broad Gaussian models.  
The obtained line widths are $\sigma$ = 44.5$^{+12.9}_{-7.3}$ eV and 
$\sigma$ = 5.0 $^{+1.1}_{-0.9}$ eV, or $v_{\rm Fe}$ = 2100 $^{+600}_{-340}$ km s$^{-1}$ and 
$v_{\rm Si} = $840 $^{+180}_{-150}$ km s$^{-1}$ in terms of the Doppler velocity dispersion, 
for the high-energy (mainly iron) and low-energy (mainly silicon) lines, respectively.  
Marshall et al.  (2002) also measured finite widths from the Doppler-shifted lines, 
although the widths were not much different from element to element. 
Figure~8 summarizes the velocity dispersion of various K$\alpha$ lines, from the present work 
and Marshall et al.  (2002), as a function of the atomic number. 
Thus, the two measurements agree very well on the low-energy lines, 
while not on the high-energy lines. 
This may be caused by the low activity of SS~433 during the observation by Marshall et al.  (2002), 
as suggested by the low values of $T_{0}$. 
If the high-energy elements, above calcium, of Marshall et al.  (2002) are excluded, figure~8 
reveals a clear positive dependence of the velocity dispersion on the atomic number. 

In figure~8, the dashed curves represent the velocity dispersion 
expected when the line widths are caused solely by thermal broadening, for several 
representative ion kinetic temperatures.  
Thus, the measured line widths are too large to be explained in terms of thermal Doppler 
effect, for a reasonable range of temperature where the atomic emission lines are 
significantly emitted. 
Furthermore, the measured positive correlation of the velocity dispersion on the atomic 
number is opposite to what is predicted by the thermal Doppler effect. 
For these reasons, the thermal broadening is concluded to be inappropriate as an 
account of the measured line widths.  

The Compton scattering is another candidate of generating the line widths. 
The X-rays lines may be narrow when produced in the jets, 
and then Compton scattered in a surrounding medium to get broadened. This may work, because the 
optical depth of the jets to the electron scattering is of the order of unity (Kotani 1998). 
In this case, a single Compton scattering in a medium of electron temperature $T_{\rm e}$ is 
expected to shift the X-ray line energy $E$ by $\Delta E \sim E (4 k T_{\rm e} - E) /  m_{\rm e} c^{2}$, 
where $m_{\rm e} c^{2}$ is the electron rest-frame energy. 
Then, the measured values of $\Delta E \sim$ 5 eV for silicon ($E \sim 1.8$ keV) and 
$ \Delta E \sim$ 45 eV for iron ($E \sim 6.7$ keV) require $T_{\rm e} \simeq $  0.8 keV and 2.5 keV, 
respectively. 
The discrepant values of $T_{\rm e}$ may be explained by a picture that the high-energy lines 
are produced in a region closer to to the central engine. However, such a scattering medium,  
having the necessary Compton optical depth ($\sim$ 1) and the inferred low $T_{\rm e}$,  
would inevitably produce prominent edges from partially ionized atoms, which are not 
seen in the spectra. Therefore, Compton scattering may not be appropriate, either. 

Another possible origin of the line widths and its dependence on the atomic number is a 
progressive collimation of the conical opening angle of the jets.  
Marshall et al.  (2002)  ascribed the line widths to the Doppler broadening that may   
result from the conical jet outflow, and concluded that 
the emission occurs in a freely expanding region of constant collimation with an 
opening half-cone angle of $\Theta$ = 1$^{\circ}$.23;  
the opening angle produces a transverse velocity of the jet, 
which causes the obtained widths. 
Assuming an oppositely directed pair of jets observed at an angle $\alpha$ to the line of sight, 
the Doppler shifts of the blue and red jets are given by 
\begin{eqnarray*}
1 + z = \gamma \{ 1 \pm \beta \ {\rm cos} (\alpha) \}, 
\end{eqnarray*}
where $\beta$ is the velocity of the jet in unit of $c$,  and 
$\gamma$ = $(1-\beta^2)^{-1/2}$. 
Solving this equation for $\gamma$ as
\begin{eqnarray*}
v_{\rm jet} = \beta \times c = \left\{ 1 - \frac{1}{(1 + z_{\rm av})^{2}}  \right\} ^{1/2}  \times c,  
\end{eqnarray*}
where  $z_{\rm av} = (z_{\rm bl}+z_{\rm rd})/2$, and utilizing the observed redshifts, 
$v_{\rm jet}$ = 0.2629 $c$ and $\alpha$ = 92$^{\circ}$.1 are obtained.  
The velocities of the perpendicular component against the direction of jet traveling 
are obtained as $v_{\rm Fe}'$  = $v_{\rm Fe}$  sin$(\alpha)$ = 2100 km s$^{-1}$ 
and $v_{\rm Si}'$  = $v_{\rm Si}$  sin$(\alpha)$ = 840 km s$^{-1}$. 
The velocity dispersion can be equated directly with $f v_{\rm jet} \tan (\Theta)$, 
where $f$ is a form factor depending on the emissivity distribution across 
the jet cross section. 
Assuming that the density is uniform through the cone's cross section 
and that the component of the velocity which is parallel to the jet axis 
is the same for all fluid elements in the slice, 
then a simple calculation gives $f$ = 0.74 (Marshall et al.  2002). 
Therefore, the present observation yields 
\begin{eqnarray*}
&&\Theta_{\rm Fe} = \tan ^{-1} \left( \displaystyle\frac{1}{f} \ \frac{v_{\rm Fe}'}{v_{\rm jet}} \right) = {2^{\circ}.1} ^{+0^{\circ}.6}_{-0^{\circ}.3}, \\
&&\Theta_{\rm Si} = \tan ^{-1} \left( \displaystyle\frac{1}{f} \ \frac{v_{\rm Si}'}{v_{\rm jet}} \right) = 0^{\circ}.8 \pm 0^{\circ}.2. 
\end{eqnarray*}
Since the Fe-K lines are thought to be produced close to the central engine while 
the Si-K lines to come from those positions of the jets which are $ \geq 10^{12}$ cm away 
from the center (Kotani 1998), the present results imply that the jet collimation is 
achieved over a distance of $ \geq 10^{12}$ cm, or $ \geq 10^{2}$ sec. 
This interpretation, if correct, provides valuable information as to the jet 
collimation mechanism. 

Since the collimation is thus inferred to occur on a scale much larger than the 
accretion disk, it is more likely to be achieved by magnetic fields, rather than by 
any funnel-shaped structure in the accretion disk. 

\subsection{Unidentified Features}

As mentioned in the last paragraph of subsection 3.3,  
the double-peaked features, which can be explained by neither  
the jet components nor the fluorescent lines from the accretion disk, 
were found around 1.86 keV.  
This energy corresponds to that of ``stationary''  Si {\sc xiii} K$\alpha$ line.  
From a careful look at the spectra in figure~6, 
similar features are also suggested around 2.01 keV,  
2.46 keV and 2.62 keV, tentatively identified with  Si {\sc xiv} K$\alpha$, 
S {\sc xv} K$\alpha$, and S {\sc xvi} K$\alpha$ lines. 
These unidentified features have not been detected for any other observation of SS~433 
using the Chandra HETGS, and their exact nature remains unknown.  

\section{Summary}\label{summary}

The analysis of the Chandra HETGS data of SS~433 acquired on the  
2001 May 12 have yielded the following results: 

\begin{enumerate}
\item The observed precessional phase of the jets was shifted by $\sim$ 20 d 
from the prediction based on Margon and Anderson (1989). 
\item The initial temperature of the blue and red jets are estimated to be 
$19.5^{+7.8}_{-9.5}$ and $17.2^{+9.0}_{-6.8}$ keV, respectively.  
\item The Doppler-shifted lines were found to be broad, 
and the corresponding velocity dispersions are 2100 $^{+600}_{-340}$ km s$^{-1}$ 
and $840 ^{+180}_{-150}$ km s$^{-1}$ for the high-energy and low-energy lines, respectively.  
\item While the observed line widths cannot be explained satisfactorily by the 
thermal Doppler effect or the Compton scattering, a successful explanation may be 
obtained by assuming that the line widths originate from the jet opening angle, 
and that the angle becomes narrower as the jets travel away from the central engine.  
\end{enumerate}

\bigskip

M. N. is supported by the Junior Research Associate Program of RIKEN.  
M. N. also would like to thank Dr. H. Matsumoto, Dr. H. L. Marshall, and the members of 
Center for Space Research (MIT)  for the study of the Chandra HETGS. 
We thank the GAO for the good optical data and the meaningful discussion and suggestion.


\begin{table}
\caption{Observation log.}\label{tab:Obslog}
\begin{tabular}{cccc}
\hline\hline\\[-10pt]
Start time (UT) & End time (UT)    & Orbital phase$^{*}$ & Exposure time\\
\hline\\[-10pt]
2001/05/12 10:38 & 2001/05/12 16:33 & 0.2832 $\pm$ 0.0094 & 19.7 ks  \\[0pt]
\hline\hline\\
\end{tabular}\\
\footnotesize
$^{*}$ Calculated from the parameters given by Gladyshev et al.  (1987).
\end{table}


\begin{table}[htb]
\begin{minipage}{0.5\textwidth}
\caption{Results of the model fit to the HETGS/HEG spectrum of SS~433 in the 4.0--9.0 keV band.$^{*}$}
\small
\begin{tabular*}{\textwidth}{@{\hspace{\tabcolsep}\extracolsep{\fill}}lcc}
\hline\hline\\[-15pt]
Parameter & Blue & Red\\
\hline\\[-15pt]
Redshift                         & 0.0460 $\pm$ 0.0013    & 0.0265$\pm$0.0014      \\[0pt]\hline
Line fluxes$^{\dagger}$        &                        &                         \\[0pt]
Fe {\sc xxv}   K$\alpha$  (6.70) & 38.5 $^{+7.4} _{-6.5}$ & 34.5 $^{+6.4} _{-6.1}$  \\[0pt]
Fe {\sc xxvi}  K$\alpha$  (6.97) & 17.6 $^{+5.5} _{-5.0}$ & 14.3 $^{+5.2} _{-4.7}$  \\[0pt]
Ni {\sc xxvii} K$\alpha$  (7.80) & 9.2  $^{+6.3} _{-5.3}$ & $<$  5.1               \\[0pt]
Fe {\sc xxv}   K$\beta$   (7.90) & 5.3  $^{+6.2} _{-5.0}$ & 4.7  $^{+6.1} _{-4.7}$  \\[0pt]
Fe {\sc i}  K$\alpha$     (6.40) & \multicolumn{2}{c}{6.4 $^{+5.2} _{-4.8}$} \\[0pt]
Fe {\sc i}  K$\beta$      (7.06) & \multicolumn{2}{c}{3.8 $^{+3.3} _{-2.6}$} \\[0pt]
\hline 
  Line width$^{\ddagger}$ &\multicolumn{2}{c}{ 39.5 $^{+7.7} _{-6.5}$  \ eV }       \\[0pt]
\hline 
  Power law $\mit\Gamma$  &\multicolumn{2}{c}{ 1.61 $\pm$ 0.07 }                            \\[0pt]
  Power law $A$       &\multicolumn{2}{c}{ ( 1.63 $\pm$ 0.03 )  $\times 10^{-2} \rm \ ph \ s^{-1} \ cm^{-2}$}\\[0pt]
\hline
  $N_{\rm H}$$^{\S}$  &\multicolumn{2}{c}{ 1.09 $\times 10^{22}$\ cm$^{-2}$  (fixed) }\\[0pt]
\hline
 $C$-statistic          &\multicolumn{2}{c}{703.57 ( d.o.f. =  594 )}       \\[0pt]
\hline\hline
\end{tabular*}
\\
\noindent
$^{*}$ All uncertainties refer to statistical 90\% confidence limits. 
  In the case of redshift, a systematic error of $\pm$ 0.0033 must be added. \\
$^{\dagger}$ The measured line fluxes, in unit of 10$^{-5}$ ph s$^{-1}$ cm$^{-2}$. 
The number in parenthesis shows the rest-frame line energy, in unit of keV. \\
$^{\ddagger}$ Common to all the ion species. \\
$^{\S}$ Column density of photoelectric absorption, determined in the soft energy band.\\
\label{tab:fitresult_high}
\end{minipage}
\end{table}


\begin{table}[htb]
\begin{minipage}{0.5\textwidth}
\caption{Results of the model fit to the HETGS / HEG $+$ MEG spectra of SS~433 in the 1.5--4.0 keV band.$^{*}$}
\small
\begin{tabular*}{\textwidth}{@{\hspace{\tabcolsep}\extracolsep{\fill}}lcc}
\hline\hline\\[-15pt]
Parameter &  Blue & Red\\
\hline\\[-15pt]
Redshift     & 0.0444 $\pm$ 0.0007 & 0.0220 $\pm$ 0.0008 \\[0pt]\hline
Line fluxes$^{\dagger}$        &                        &                         \\[0pt]
Si {\sc xiii} $f$  K$\alpha$ $^{\|}$  (1.85) & 1.0 $^{+1.2} _{-1.0}$ & $<$ 0.9                \\[0pt]
Si {\sc xiii}      K$\alpha$          (1.86) & 2.7 $^{+1.3} _{-1.2}$ & 4.2 $^{+1.3} _{-1.3}$  \\[0pt]
Si {\sc xiv}       K$\alpha$          (2.01) & 7.3 $^{+1.6} _{-1.5}$ & 5.4 $^{+1.5} _{-1.4}$  \\[0pt]
Si {\sc xiv}        K$\beta$          (2.38) & 3.0 $^{+2.4} _{-2.1}$ & $<$ 1.7                \\[0pt]
S  {\sc xv} $f$    K$\alpha$ $^{\|}$  (2.45) & 1.9 $^{+2.4} _{-1.9}$ & $<$ 1.7                \\[0pt]
S  {\sc xv}        K$\alpha$          (2.46) & 3.2 $^{+2.3} _{-2.0}$ & 2.9 $^{+2.2} _{-2.0}$  \\[0pt]
S  {\sc xvi}       K$\alpha$          (2.62) & 4.4 $^{+2.1} _{-1.9}$ & 2.3 $^{+2.0} _{-1.7}$  \\[0pt]
\hline 
  Line width$^{\ddagger}$ &\multicolumn{2}{c}{ 5.2 $^{+1.1} _{-0.9}$  \ eV }       \\[0pt]
\hline 
  Power law $\mit\Gamma$  &\multicolumn{2}{c}{ 1.21 $\pm$ 0.05 }                            \\[0pt]
  Power law $A$       &\multicolumn{2}{c}{ ( 9.41 $\pm$ 0.07 )  $\times 10^{-3} \rm \ ph \ s^{-1} \ cm^{-2}$}\\[0pt]
\hline
  $N_{\rm H}$         &\multicolumn{2}{c}{ ( 1.09 $\pm$ 0.07 ) $\times 10^{22}$\ cm$^{-2}$  }\\[0pt]
\hline
 $C$-statistic          &\multicolumn{2}{c}{1468.32 ( d.o.f. =  1235 )}       \\[0pt]
\hline\hline
\end{tabular*}
\\
\noindent
$^{*}$ All uncertainties refer to statistical 90\% confidence limits. 
  In the case of redshift, a systematic error of $\pm$ 0.0017 must be added. \\
$^{\|}$ Forbidden lines.\\
The other annotation symbols are the same as in table~2. 
\label{tab:fitresult_low}
\end{minipage}
\end{table}


\begin{table}[htb]
\begin{minipage}{0.5\textwidth}
\caption{Results of the model fit to the HETGS / HEG $+$ MEG spectra of SS~433 in the 1.5--9.0 keV band.$^{*}$}
\small
\begin{tabular*}{\textwidth}{@{\hspace{\tabcolsep}\extracolsep{\fill}}lcc}
\hline\hline\\[-15pt]
Parameter &  Blue & Red\\
\hline\\[-15pt]
Redshift          & 0.0443 $\pm$ 0.0006 & 0.0227 $\pm$ 0.0007 \\[0pt]\hline
Line fluxes$^{\dagger}$        &                        &                         \\[0pt]
Si {\sc xiii} $f$  K$\alpha$ $^{\|}$     (1.85) & 1.0 $^{+1.3} _{-1.0}$ & $<$ 0.8  \\[0pt]
Si {\sc xiii}      K$\alpha$             (1.86) & 2.9 $^{+1.4} _{-1.4}$ & 4.3 $^{+1.5} _{-1.4}$  \\[0pt]
Si {\sc xiv}       K$\alpha$             (2.01) & 7.8 $^{+1.8} _{-1.8}$ & 5.8 $^{+1.6} _{-1.5}$  \\[0pt]
Si {\sc xiv}        K$\beta$             (2.38) & 3.0 $^{+2.5} _{-2.2}$ & $<$ 1.5  \\[0pt]
S  {\sc xv} $f$    K$\alpha$ $^{\|}$     (2.45) & 2.0 $^{+2.4} _{-2.0}$ & $<$ 1.5  \\[0pt]
S  {\sc xv}        K$\alpha$             (2.46) & 3.1 $^{+2.4} _{-2.1}$ & 2.8 $^{+2.3} _{-2.0}$  \\[0pt]
S  {\sc xvi}       K$\alpha$             (2.62) & 4.5 $^{+2.2} _{-2.0}$ & 2.4 $^{+2.1} _{-1.8}$  \\[0pt]
\hline 
  Line width$^{\#}$ &\multicolumn{2}{c}{ 5.0 $^{+1.1} _{-0.9}$  \ eV }       \\[0pt]
\hline 
Fe {\sc xxv}   K$\alpha$   (6.70) & 42.6 $^{+7.3} _{-6.9}$ & 30.9 $^{+6.6} _{-6.1}$  \\[0pt]
Fe {\sc xxvi}  K$\alpha$   (6.97) & 16.9 $^{+5.8} _{-5.2}$ & 10.9 $^{+5.2} _{-4.6}$  \\[0pt]
Ni {\sc xxvii} K$\alpha$   (7.80) & 9.6 $^{+6.7} _{-5.6}$ & $<$ 5.9  \\[0pt]
Fe {\sc xxv}   K$\beta$    (7.90) & 3.4 $^{+6.3} _{-3.4}$ & 3.1 $^{+6.3} _{-3.1}$  \\[0pt]
Fe {\sc i}  K$\alpha$      (6.40) & \multicolumn{2}{c}{6.8 $^{+4.9} _{-4.5}$} \\[0pt]
Fe {\sc i}  K$\beta$       (7.06) & \multicolumn{2}{c}{3.4 $^{+3.3} _{-2.6}$} \\[0pt]
\hline 
  Line Width$^{**}$ &\multicolumn{2}{c}{ 44.5 $^{+12.9} _{-7.3}$  \ eV }       \\[0pt]
\hline 
  Power law $\mit\Gamma$  &\multicolumn{2}{c}{ 1.40 $\pm$ 0.04 }                            \\[0pt]
  Power law $A$       &\multicolumn{2}{c}{ ( 1.18 $\pm$ 0.07 ) $\times 10^{-2} \rm \ ph \ s^{-1} \ cm^{-2}$}\\[0pt]
\hline
  $N_{\rm H}$         &\multicolumn{2}{c}{ ( 1.31 $\pm$ 0.06 ) $\times 10^{22}$\ cm$^{-2}$   }\\[0pt]
\hline
 $C$-statistic          &\multicolumn{2}{c}{ 2311.75 ( d.o.f. = 1964 )}       \\[0pt]
\hline\hline
\end{tabular*}
\\
\noindent
$^{*}$ All uncertainties refer to statistical 90\% confidence limits. 
  In the case of redshift, a systematic error of $\pm$ 0.0017 must be added. \\
$^{\#}$ The width for low-energy lines.\\
$^{**}$ The width for high-energy lines.\\
The other annotation symbols are the same as table~2. 
\label{tab:fitresult_all}
\end{minipage}
\end{table}


\begin{figure}
  \begin{center}
    \FigureFile(100mm,100mm){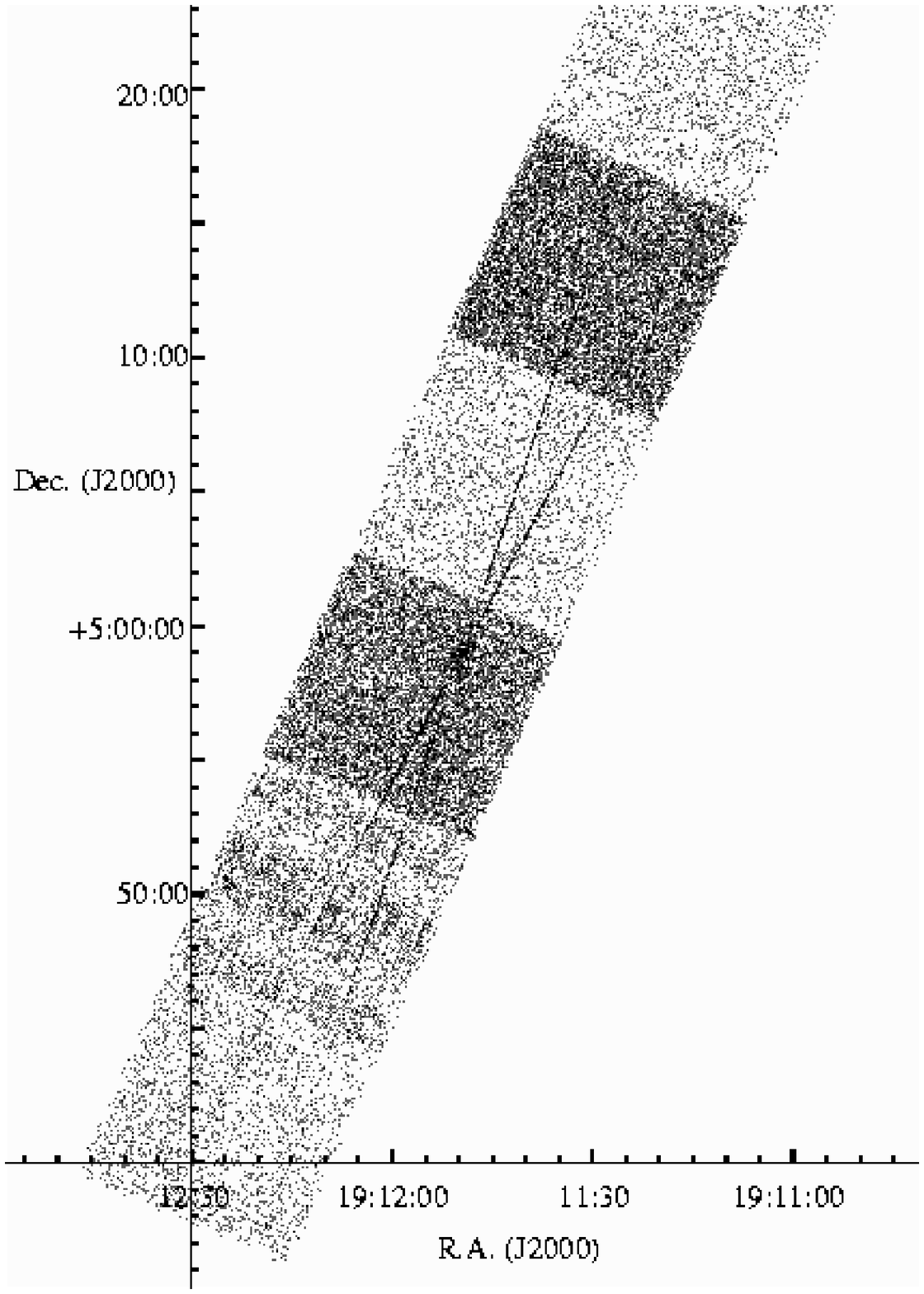}
  \end{center}
  \caption{Zeroth-order image and dispersion lines of SS~433 taken with the 
ACIS-CCD on 2001 May 12.  The central point source (SS~433) and 4 lines 
(grating spectra) are shown in J2000 coordinates.}
\end{figure}


\begin{figure}
  \begin{center}
    \FigureFile(100mm,100mm){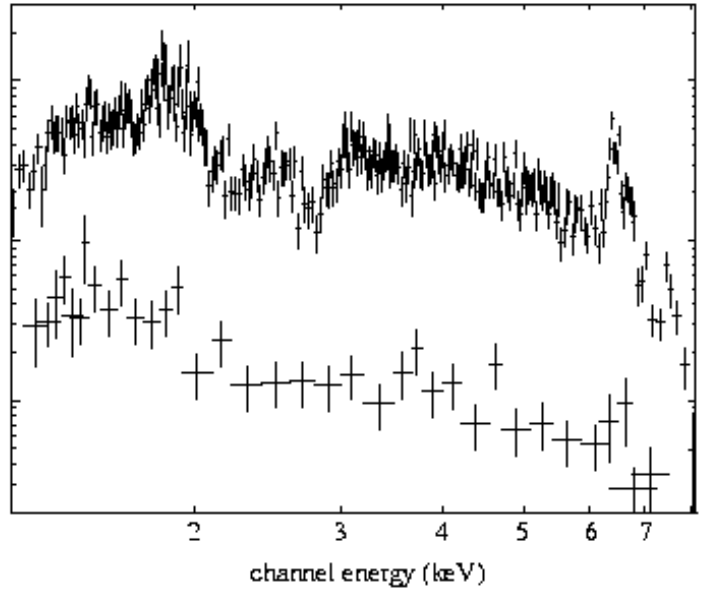}
  \end{center}
  \caption{Chandra HETGS spectra of SS~433 acquired on 2001 May 12, 
shown for the total (1.0--8.0 keV) band. 
The upper and lower spectra correspond to the source and background, respectively. }
\end{figure}


\begin{figure}
  \begin{center}
    \FigureFile(100mm,100mm){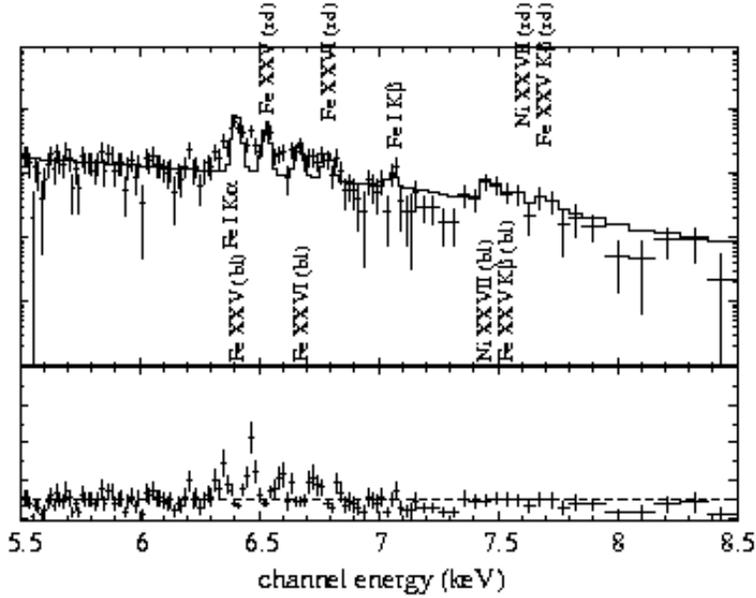}
  \end{center}
  \caption{Same on-source spectrum as presented in figure~2, shown over the iron K-line energy band. 
The solid histograms represent the best-fit jet emission model, in which the jet redshifts are 
left free but the lines are constrained to be narrow. 
The location of emission lines are labeled with ion species 
and the jet ID; ``bl'' denotes lines from the ``blue jet'',  
and ``rd'' denotes those from the ``red jet''.  
The bottom panel shows the intensity ratio against the best-fit model.}
\end{figure}


\begin{figure}
  \begin{center}
    \FigureFile(100mm,100mm){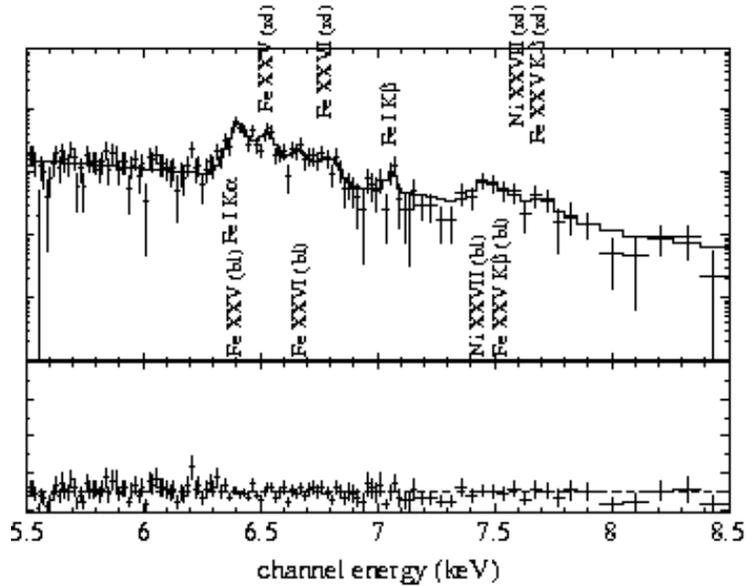}
  \end{center}
  \caption{Same as figure~3 , 
but the Doppler-shifted emission lines in the model are 
allowed to have a common width.}
\end{figure}


\begin{figure}
  \begin{center}
    \FigureFile(100mm,100mm){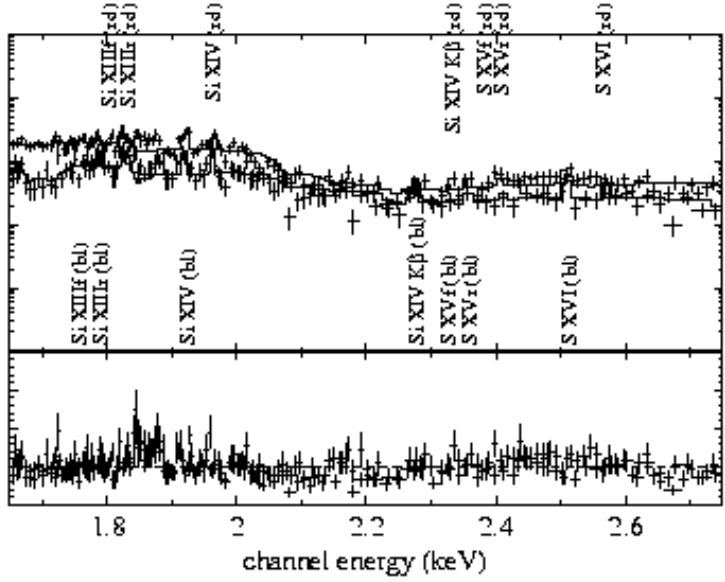}
  \end{center}
  \caption{Chandra HEG (lower) and MEG (upper) spectra of SS~433 taken on 2001 May 12, 
shown over the silicon K-line energy band. 
The subscripts ``r'' and ``f'' are for the resonance and forbidden components, respectively.
The Gaussian lines are constrained to be narrow. The bottom panel shows the ratio against the 
best-fit model.}
\end{figure}

\begin{figure}
  \begin{center}
    \FigureFile(100mm,100mm){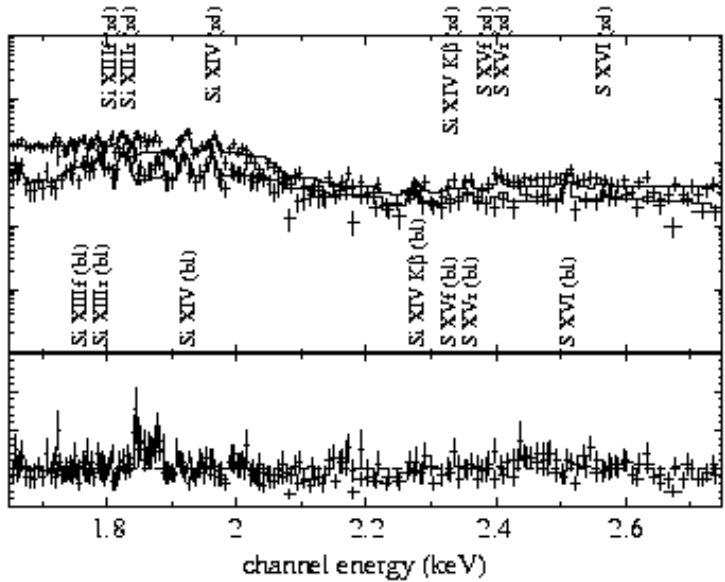}
  \end{center}
  \caption{Same as figure~5, 
but the emission lines in the model are allowed to have a common width.}
\end{figure}


\begin{figure}
  \begin{center}
    \FigureFile(100mm,100mm){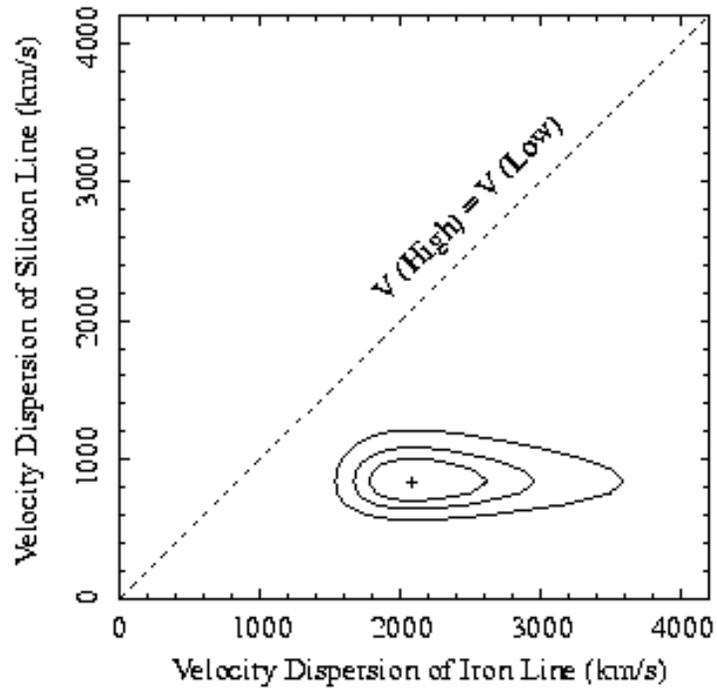}
  \end{center}
  \caption{Confidence contours for the velocity dispersion of Fe {\sc xxv} K$\alpha$ 
and Si {\sc xiii} K$\alpha$ lines. The contours correspond to the 99\%, 90\% and 68\% 
confidence levels from outside to inside. }
\end{figure}


\begin{figure}
  \begin{center}
    \FigureFile(100mm,100mm){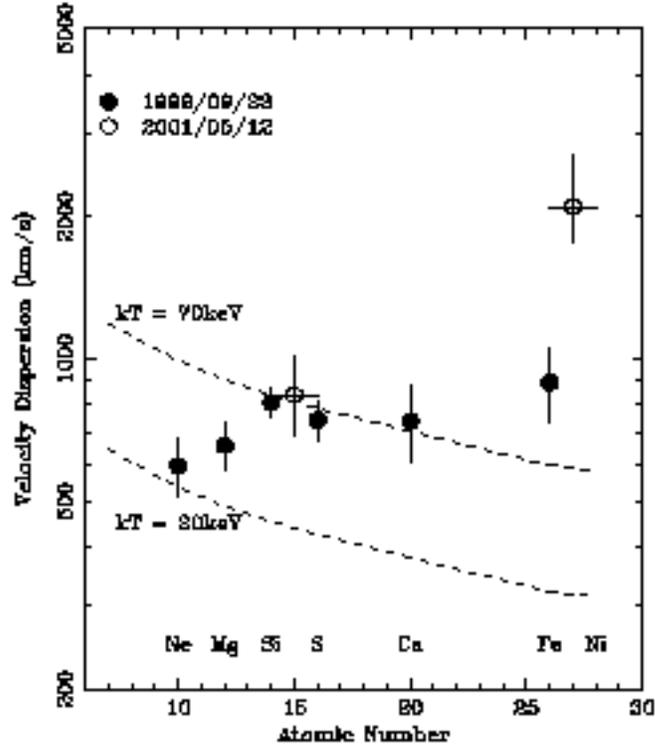}
  \end{center}
  \caption{Atomic number of K$\alpha$ line vs. its velocity dispersion.  
The filled and open circles correspond to the Chandra HETGS data observed on 
1999 September 23 (Marshall et al.  2002) and 2001 May 12 (the present work), 
respectively.  
The dashed lines are velocity dispersions of the K$\alpha$ lines expected for 
thermal Doppler broadening. 
The upper one is for an ion kinetic temperature of 70 keV, and lower is for 20 keV. }
\end{figure}

\end{document}